\documentclass[prl,twocolumn,superscriptaddress,amsmath,amssymb,showpacs,longbibliography]{revtex4-1}
\usepackage{hyperref}
\pdfoutput=1
\usepackage{graphicx}

\usepackage{amsmath}
\usepackage{bm}
\usepackage{amssymb}
\usepackage{textcomp}
\usepackage{stmaryrd}
\usepackage{ulem}
\usepackage{gensymb}
\usepackage{units}

\usepackage{import}
\usepackage{color}
\usepackage{verbatim}

\usepackage{float}

\usepackage{sistyle} 

\def\XXint#1#2#3{{\setbox0=\hbox{$#1{#2#3}{\int}$}
     \vcenter{\hbox{$#2#3$}}\kern-.5\wd0}}

\begin{document}

\title{Dipolar and magnetic properties of strongly absorbing hybrid interlayer excitons in pristine bilayer MoS$_2$}

\author{Etienne Lorchat}
\affiliation{Technische Physik, Wilhelm-Conrad-R\"ontgen-Research Center for Complex Material Systems,
Universität W\"urzburg, Am Hubland, D-97074 Würzburg, Germany}
\author{Malte Selig}
\affiliation{Technische Universit\"at Berlin, Germany}
\author{Florian Katsch}
\affiliation{Technische Universit\"at Berlin, Germany}
\author{Kentaro Yumigeta}
\affiliation{School for Engineering of Matter, Transport, and Energy, Arizona State University, Tempe, Arizona 85287, USA}
\author{Sefaattin Tongay}
\affiliation{School for Engineering of Matter, Transport, and Energy, Arizona State University, Tempe, Arizona 85287, USA}
\author{Andreas Knorr}
\affiliation{Technische Universit\"at Berlin, Germany}
\author{Christian Schneider}
\affiliation{Technische Physik, Wilhelm-Conrad-R\"ontgen-Research Center for Complex Material Systems,
Universität W\"urzburg, Am Hubland, D-97074 W\"urzburg, Germany}
\author{Sven H\"ofling}
\affiliation{Technische Physik, Wilhelm-Conrad-R\"ontgen-Research Center for Complex Material Systems,
Universität W\"urzburg, Am Hubland, D-97074 W\"urzburg, Germany}

\begin{abstract}

Van der Waals heterostructures composed of transition metal dichalcogenide monolayers (TMDs) are characterized by their truly rich excitonic properties which are determined by their structural, geometric and electronic properties: In contrast to pure monolayers, electrons and holes can be hosted in different materials, resulting in highly tunable dipolar manyparticle complexes. However, for genuine spatially indirect excitons, the dipolar nature is usually accompanied by a notable quenching of the exciton oscillator strength. Via electric and magnetic field dependent measurements, we demonstrate, that a slightly biased pristine bilayer MoS$_2$ hosts strongly dipolar excitons, which preserve a strong oscillator strength. We scrutinize their giant dipole moment, and shed further light on their orbital- and valley physics via bias-dependent magnetic field measurements.   

\end{abstract}

\maketitle

\paragraph*{Introduction}

The interest in excitons hosted in atomically thin materials was initially sparked by their giant oscillator strength and large binding energies, resulting from the canonical interplay of reduced dielectric screening and confinement of charge carriers in an atomically thin sheet \cite{wang2018}. The spectacular finding of extreme optical activity triggered a plethora of experiments, analyzing their chirality \cite{schaibley2016valleytronics}, their magnetic behavior \cite{gong2013magnetoelectric, srivastava2015valley, stier2016exciton, aivazian2015magnetic,robert2020measurement}, interactions yielding higher order multi-particle complexes \cite{Chernikov2014} and finally driving the field of opto-electronic applications \cite{wang2012electronics}. In the latter, in particular the coupling to confined microcavity modes became a field of particular interest, giving rise to pronounced polaritonic phenomena up to ambient conditions \cite{lundt2016room, liu2015strong, schneider2018two}. More recently, the sheet nature of TMDCs, which allows almost arbitrary stacking and alignment of multiple monolayers was harnessed to compose more complex electronic structures. One recurrent scheme in devices based on such van der Waals heterostructures are charge transfer phenomena, which yield the formation of dipolar excitons with electrons and holes confined in different layers \cite{rivera2015observation}. While such excitons do have appealing properties, including enhanced nonlinearities borrowed from their dipolar character, the giant oscillator strength, which initially sparked the success of TMD excitons is strongly compromised by the reduced spatial overlap of electron and holes in separate layers. However, recently it was suggested that the hybridization of hole states in a pristine bilayer of MoS$_2$ enables the formation of dipolar excitons, which combine a permanent dipole moment with a giant oscillator strength \cite{gerber2019interlayer}. Here, we give evidence for this new species of TMD exciton: The dipolar character is clearly evidenced in field-dependent absorption measurements, while maintaining a substantial optical absorption. We further scrutinize the orbital- and valley composition by assessing the excitonic g-factor  in the presence of a static electric field, and develop a consistent microscopic theory describing our main findings.  

\paragraph*{Experimental results}

\begin{figure}[!th]
\begin{center}
\includegraphics[width=0.97\linewidth]{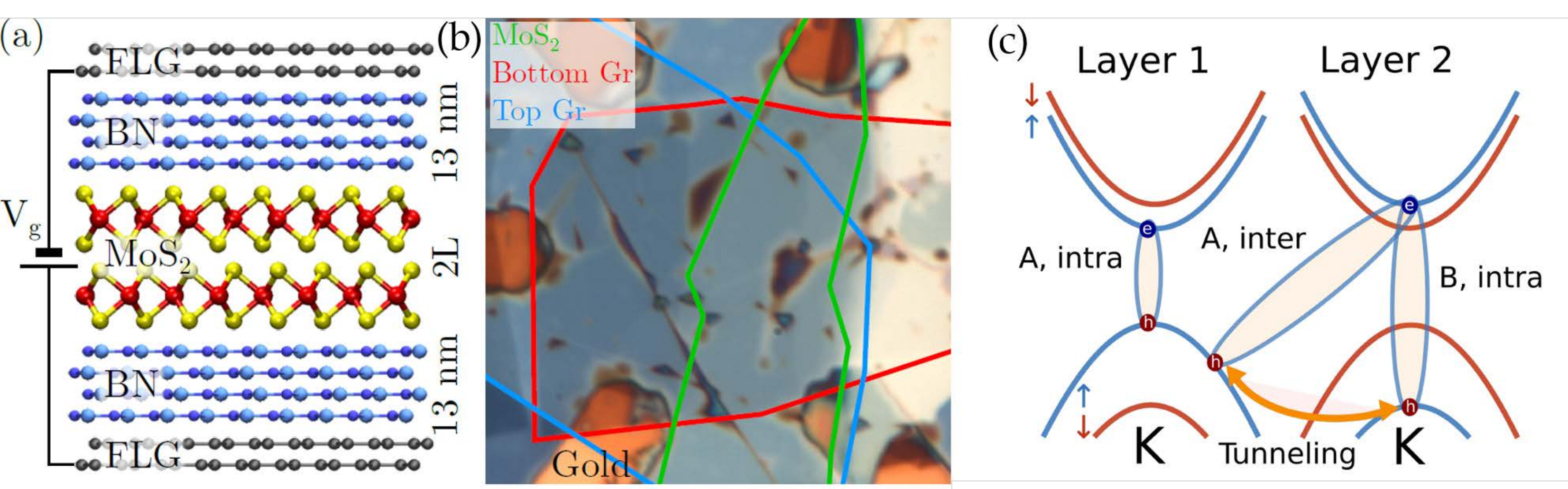}
\caption{a) Sketch of the device, composed of a few layer graphene, boron nitrite, bilayer MoS$_2$,boron nitrite, few layer graphene sandwhich. b) Optical micrograph of the sample. c) Sketch of the exciton states relevant for the optical response in bilayer MoS$_2$ at the K point. At the K' point (not shown), the spin bands are reversed. }

\label{Fig1b}
\end{center}
\end{figure}

In order to characterize interlayer excitons with strong dipole moment in pristine MoS$_2$ TMD bilayers, we fabricate a van der Waals heterostructure composed of few layer graphene (FLG), an approx. 13 nm thick layer of hexagonal boron nitride (BN), a pristine bilayer MoS$_2$, another 13 nm thick BN and FLG (Fig. \ref{Fig1b} a). Both FLG are electrically contacted to achieve a top and backgate which allows us to apply an out of plane electric field. A microscope image of the fully assembled device is shown in Fig. \ref{Fig1b} b). 
The excitonic optical response of bilayer MoS$_2$ is composed by the so-called A- and B- exciton transitions at the K and K' point, respectively. As opposed to a monolayer, charge transfer processes further allow for the emergence of spatially indirect excitonic complexes in bilayers, which enriches the optical response. As schematically captured in Fig. \ref{Fig1b} c) and anticipated in \cite{gerber2019interlayer}, a hybrid resonance evolves by hybridizing an A-type interlayer transition with a B-type intralayer resonance. Such a hybrid transition consequently admixes the dipolar properties of interlayer excitons with the strong oscillator strength of TMDC intralayer resonances.

\begin{figure}[!th]
\begin{center}
\includegraphics[width=0.97\linewidth]{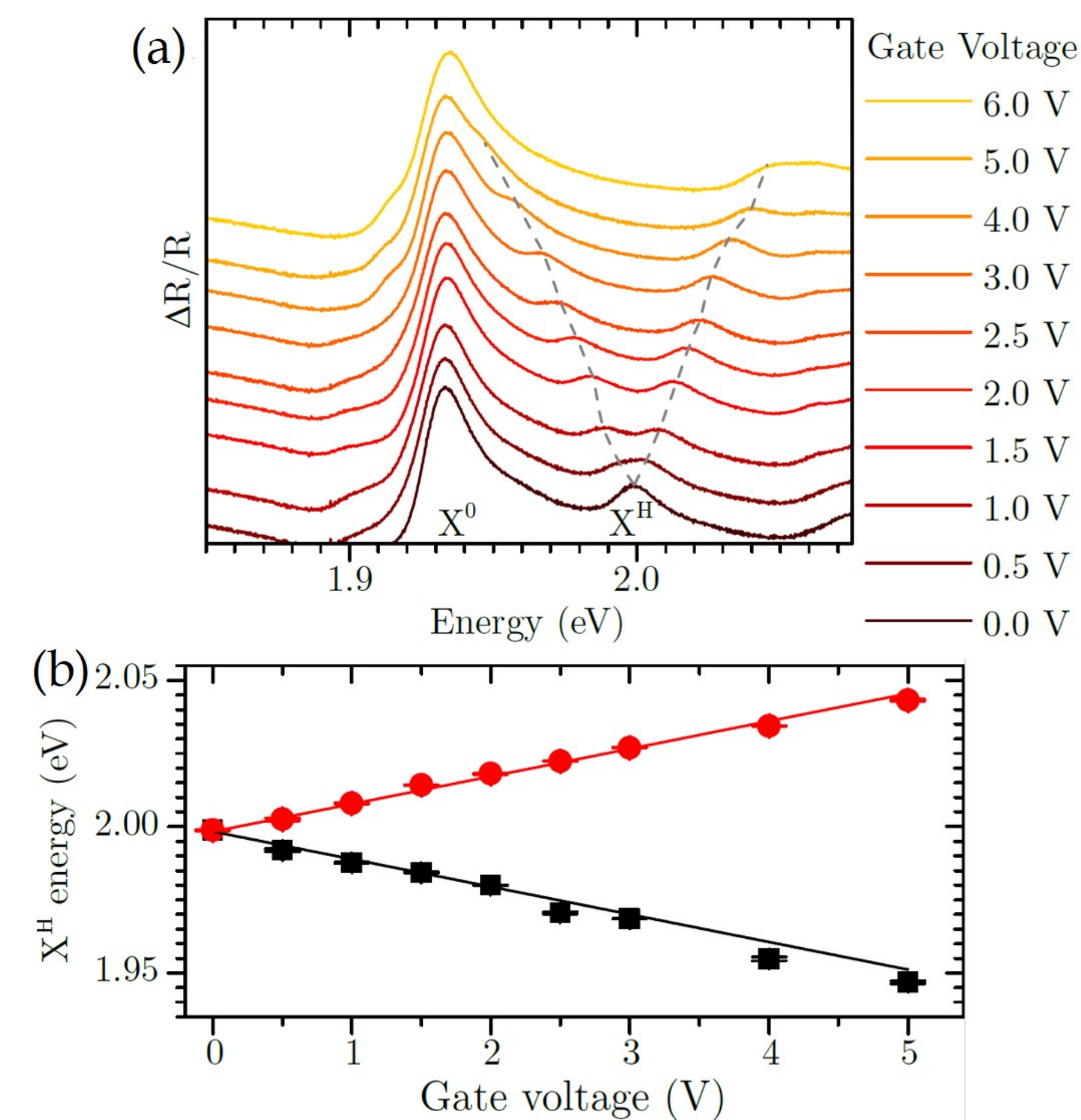}
\caption{a) Cascade plot of the sample differential reflectivity at various gate voltage. b) X$^{\mathrm{H}}$ energy with respect to the gate voltage for both dipole (black squares and red dot). The red and black lines are linear fits to the data to extract the dipole size.}
\label{Fig2}
\end{center}
\end{figure}

We first probe the optical response of our device via white light reflection spectroscopy at cryogenic temperatures (4 K), and plot the differential reflectivity $\frac{\Delta R}{R}$ normalized with respect to a reference spectrum recorded next to the structure. At zero gate voltage, we retrieve the characteristic absorption spectrum of bilayer MoS$_2$, which is composed of two significant optical resonances in the energy range between 1.8 and 2.1 eV. These two absorption peaks have been assigned to the neutral A exciton transition at 1.95 eV, as well as the hybrid excitonic state which mixes A-inter- and B-intralayer states (see (Fig. \ref{Fig1b} c)  (denoted as X$^{\mathrm{H}}$ in the following). We point out, that this hybrid mode displays a significant reflection signal, caused by a strong oscillator strength which is only smaller by approx. a factor of 4 as compared to the intralayer A-exciton. 

Evident from the bias dependent reflectivity measurements in Fig. \ref{Fig2} a), the applied static electric field  couples to out of plane dipoles via the Stark effect, modifying the exciton energy as $\Delta \mathrm{E_X} = \mathbf{d} \cdot \mathbf{E}$ with $\mathbf{d}$ the out of plane electric dipole and $\mathbf{E}$ the electric field. Since the X$^{\mathrm{H}}$ dipole is not polarized in our white light reflection experiment, both dipole (pointing from layer 1 to layer 2 and vice versa) coexist at the same energy at zero electric field. The out of plane electrical field lifts this degeneracy ending up with two different absorption lines in the spectrum (Fig. \ref{Fig2} a). We note, that the application of only 5 V to our device yields a giant dipolar splitting of 100 meV of the two modes, which is only accompanied by a modest quenching of the oscillator strength.

For a quantitative analysis, we estimate the electric field from the gate voltage and the device geometry via a plate capacitor model ($\varepsilon_{\mathrm{BN}} = 3.67$, $\varepsilon_{\mathrm{MoS2}} = 6.8 $ \cite{laturia2018dielectric}, following the approach discussed in \cite{wang2018electrical} (see supplementary notes).  We extracted the emission energy of both X$^{\mathrm{H}}$ upper and lower branch which scales linearly with the gate voltage. By fitting our data, we directly yield a giant dipole moment of 0.48 $\pm0.1$ nm  for X$^{\mathrm{H}}$, which is approx. a factor of 2 smaller than the interlayer distance of 1 nm (Fig. \ref{Fig2} b), which matches the phenomenological expectation of a layer-localized electron coupling to a hole which is delocalized over both layers.  
 
 \begin{figure}[!th]
\begin{center}
\includegraphics[width=0.97\linewidth]{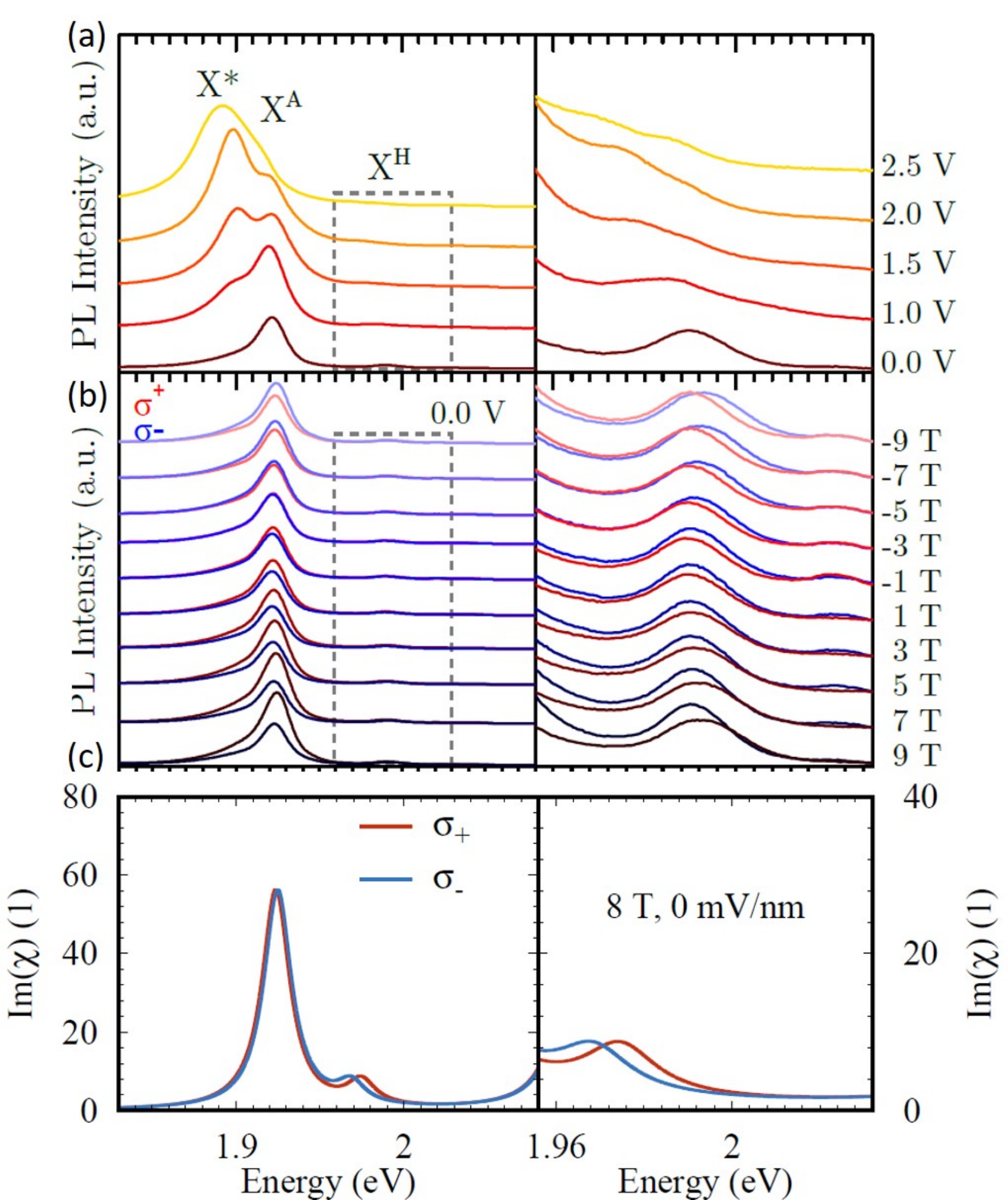}
\caption{a) Left panel, PL spectra at different gate voltage at 0 magnetic field the lowest lying emission line correspond to the trion emission and is denoted X$^{\mathrm{*}}$, the neutral intralayer exciton is denoted X$^{\mathrm{A}}$ and the hybrid exciton is denoted X$^{\mathrm{H}}$. Right panel, close up around X$^{\mathrm{H}}$ b) PL spectra at 0 gate voltage at different magnetic field. right panel is a close up around X$^{\mathrm{H}}$. c)Polarization resolved dielectric susceptibility at a magnetic field of $B_z=8 T$.}
\label{Fig3}
\end{center}
\end{figure}

To further analyze the character of the hybrid excitonic mode, we study its behavior in an externally applied magnetic field. Therefore, we excite our bilayer non-resonantly with a linearly polarized 532 nm cw laser (5 mW focused onto a 3 µm diameter spot) and record the emitted photoluminescence (PL) for varying gate voltage and applied magnetic fields. We furthermore applied polarization resolved spectroscopy to extract the characteristic Zeeman-splitting which is reflected by the energetic difference of the circular left and right polarized emission. 

Fig. \ref{Fig3} a) depicts the evolution of the photoluminescence as a function of the applied gate voltage at 0T. Due to the lower energy of the X$^{\mathrm{A}}$ transition, most of the luminescence stems from this mode. The PL from X$^{\mathrm{H}}$ is quenched by approximately a factor of 10, but remains visible in the spectra (see right, zoomed panel). Evidently, when the bias voltage is raised up to 2.5 V, a third, red-shifted peak appears and eventually dominates the luminescence response the spectrum, which we assign to the charged intralayer excitonic state of the bilayer. 

Importantly, as depicted in Fig. \ref{Fig3} b), a notable polarization splitting in both, the  X$^{\mathrm{A}}$ transition as well as the  X$^{\mathrm{H}}$ transition occurs in the presence of an applied magnetic field (here, no electric field is applied). A closer inspection indeed reveals, that this magnetic field splitting of the two modes displays an opposite sign, and a substantially modified magnitude: at -9T, the X$^{\mathrm{A}}$ mode experiences a splitting of ~0.8 meV, whereas the X$^{\mathrm{H}}$ mode splitting exceeds -2 meV. To further quantify this behavior, we analyzed the extracted peak positions as a function of the applied magnetic field, yielding the g-factor of the X$^{\mathrm{A}}$ exciton as -2. This, indeed, contrasts the commonly observed g-factors fluctuationg around -4 for A-excitons in TMDC monolayers, but we note that in particular in MoS$_2$ monolayers, g-Factors of -2 have been reported \cite{goryca2019revealing, cadiz2017excitonic}. In our experiment, the g-factor of the hybrid X$^{\mathrm{H}}$ , however, acquires a value of  approx. 4.2. This already suggests that X$^{\mathrm{A}}$ and X$^{\mathrm{H}}$ are of fundamentally different character, and admix different valley contributions.  

To investigate the Zeeman and Stark shifts of intra- and interlayer states more carefully, we carry out a microscopic analysis in the Heisenberg equation of motion framework. Starting point is the electronic bilayer Hamiltonian
\begin{align}
H &= \sum_{l \lambda \xi s \mathbf{k}} E^{l \lambda \xi s}_\mathbf{k} \lambda^{\dagger l \xi s}_\mathbf{k}  \lambda^{l \xi s}_\mathbf{k} + \sum_{l s \xi \lambda \mathbf{k}} t^{l \bar{l} \xi s \lambda}_\mathbf{k} \lambda^{l \xi s}_\mathbf{k} \lambda^{\bar{l} \xi s}_\mathbf{k} \nonumber \\ 
&+ \sum_{l \xi s \lambda \mathbf{k}} \mathbf{d}^{l l \lambda \xi \bar{\lambda}}_\mathbf{k} \cdot  \mathbf{E}^{opt} \lambda^{\dagger l \xi s}_\mathbf{k} \bar{\lambda}^{l' \xi s }_\mathbf{k} \nonumber \\ &+ \sum_{ll'\xi\xi'ss'\lambda, \lambda', \mathbf{k},\mathbf{k'},\mathbf{q}} V^{ll'}_\mathbf{q} \lambda^{\dagger l \xi s}_\mathbf{k+q} \lambda^{l'\xi's'}_\mathbf{k'-q} \lambda^{l'\xi's'}_\mathbf{k'} \lambda^{l\xi s}_\mathbf{k}\nonumber \\
&+\sum_{l \lambda s \mathbf{k}} d^{l}_z E_z \lambda^{\dagger l \xi s}_\mathbf{k} \lambda^{l \xi s}_\mathbf{k} + \sum_{l \lambda s \mathbf{k}}\mu^{l \lambda \xi s } B_z \lambda^{\dagger l \xi s}_\mathbf{k} \lambda^{l \xi s}_\mathbf{k}\label{Hamiltonian}
\end{align}

The first term describes the dispersion of carriers in each monolayer constituting the bilayer. Here, $l=1,2$ accounts for the layer index, $\lambda=v,c$ is for the band index, $\xi=K,K'$ denotes the valley, $s=\uparrow,\downarrow$ denotes the spin and $\mathbf{k}$ the momentum. The dispersion $E^{l \lambda \xi s}_\mathbf{k}$ obeys the following symmetries 
\begin{equation}
E^{l \lambda \xi s}_{\mathbf{k}} = E^{\bar{l} \lambda \xi \bar{s}}_{\mathbf{k}},\, E^{l \lambda \xi s}_{\mathbf{k}} = E^{l \lambda \bar{\xi} \bar{s}}_{\mathbf{k}}.
\end{equation}
Here the notation $\bar{\xi}$ accounts for the opposite valley to $\xi$, i.e. $\bar{K}=K'$. Similarly, $\bar{\lambda}$ accounts for the opposite band to $\lambda$, i.e. $\bar{c}=v$.
We parametrize the electronic band structure of both individual monolayers by DFT calculations for monolayers\cite{Kormanyos2015}. The second term term accounts for the tunneling of carriers between both layers $l,\bar{l}$, which is negligible for conduction band electrons but significant for the valence band, $t^{l \bar{l} \xi s v}=$\unit[38]{meV}\cite{deilmann2018interlayer}.
The second line of the Hamiltonian accounts for light matter coupling. We assume a field propagation perpendicular to the bilayer, i.e. $\mathbf{E}^{opt}\cdot \mathbf{e}_z = 0$ (only in-plane polarized excitons).  We assume vanishing in-plane components of the interlayer dipole moment $\mathbf{d}^{l \bar{l} \lambda \bar{\lambda}}_\mathbf{k}\cdot \mathbf{e}_\parallel = 0$. Radiative coupling of intralayer excitons is taken into account via solving Maxwells equations in a planar geometry\cite{stroucken1996coherent}.
The third line in equation \ref{Hamiltonian} accounts for the Coulomb coupling of the carriers. We neglect conduction band wavefunction overlap between the layers, i.e. we disregard Dexter coupling. The Coulomb coupling describes the formation interlayer excitons $\langle v^{\dagger l} c^{\bar{l}} \rangle$. The appearing Coulomb matrix element is given by an analytic solution of the static Poisson equation\cite{ovesen2019interlayer} incorporating the dielectric environment of the sample (2 layers embedded in hBN, $\epsilon_{\text{substrate}}=4,5$). It reads
\begin{equation}
V_q^{ll'} = \frac{e^2}{\epsilon_0 \epsilon^{ll'} (q) q}
\end{equation}
with the dielectric function $\epsilon^{ll'} (q)$ which accounts for Coulomb coupling within the layers $l=l'$ or among different layers $l \neq l'$\cite{ovesen2019interlayer}.

The first term in the fourth line of equation \ref{Hamiltonian} accounts for Stark shifts of the electronic bands in the vicinity of a static out-of-plane electric field. For moderate electric fields (up to \unit[1]{V/nm}), the shifts are linear as a function of the electric field, with a slope of $\alpha=$\unit[0.05]{e nm}\cite{deilmann2018interlayer}. Bands within the same layer shift in the same direction, but the bands of the different layer shift in opposite direction\cite{deilmann2018interlayer}.
The last term accounts for the Zeeman shifts of the electronic states in an out-of-plane magnetic field. To the total magnetic moment $\mu^{l \lambda \xi s}$ three effects contribute\cite{macneill2015breaking}
\begin{equation}
\mu^{l \lambda \xi s} = \mu_{Spin}^{l s}+ \mu_{Orbital}^{l v \xi} + \mu_{Valley}^{l \lambda \xi},
\end{equation}
namely the spin contribution, the orbital contribution and the valley contribution: 

\begin{itemize}
\item the spin magnetic moment $\mu_{Spin}^{s} = 2 s \mu_B,\,s=+\frac{1}{2}/-\frac{1}{2}$ for $\uparrow/\downarrow$\cite{gong2013magnetoelectric,van2018strong}: For optically allowed transitions the contributions of valence and conduction band cancel. 
\item The orbital magnetic moment $\mu_{Orbital}^{v \xi} = 2 \xi \mu_B,\,\xi=+1/-1$ \cite{aivazian2015magnetic,srivastava2015valley}: In the monolayer, this contribution is responsible for the observed Zeeman shifts\cite{macneill2015breaking}. 
\item The valley magnetic moment $\mu_{Valley}^{\lambda \xi} = \xi \frac{m_0}{m_\lambda}\mu_B,\,\xi=+/-$ \cite{yao2008valley,xu2014spin}: Since valence and conduction bands shift in the same direction, this contribution cancels almost for intralayer excitons with similar electron and holes masses.
\item The bilayer in AB stacking exhibits inversion symmetry. This implies, that the $K$ point in layer 1 underlies the same Zeeman shifts as the $K'$ point in layer 2 and vice versa $\mu^{l \lambda \xi s} = \mu^{\bar{l} \lambda \bar{\xi} s}$.
\end{itemize}

Next, we calculate the excitonic binding energies and wavefunctions for intra- and interlayer excitons by exploiting the Wannier equation and access the linear optical response via calculating the equation of motion for the exciton, as detailed in the supplementary material. Our calculations yield binding energies of \unit[163]{meV} for intralayer and an encouragingly large binding energy of \unit[112]{meV} for the interlayer A excitons. 

Figure \ref{Fig3} illustrates the imaginary part of the dielectric susceptibility of the bilayer. Without the inclusion of the tunneling in the equation of motion (supplementary informations, equation 8) only the $A$ transition at \unit[0]{eV} and $B$ transitions occur in the spectrum (see supplementary materials). Inclusion of the tunneling yields a weak resonance above the $A$ exciton, being associated with the interlayer $A$ exciton, forming an optically allowed transition through hybridization of intralayer $B$ and interlayer $A$ excitons.

Applying a magnetic field to our model in Figure \ref{Fig3} illustrates the polarization resolved dielectric susceptibility. Here, we consider a magnetic field of \unit[8]{T}. We observe a splitting of all resonances, in particular the intralayer transitons exhibit a negative $g$ factor whereas the interlayer transition exhibits a positive one of much larger magnitude and opposite sign, clearly reproducing the phenomenological behavior captured in our experiment. A detailed analysis shows, that for the intralayer exciton, the valley contributions to the Zeeman shift for electron and hole almost cancel, and the total Zeeman shift is dictated
by the orbital contribution. In contrast, for the interlayer transition, while electron and hole stem from
different layers, the valley contribution is adding up. Additionally, while the hole is located
in the other layer compared to the intralayer A transition, the orbital contribution has
opposite sign, which adds up with the valley contribution.

\begin{figure}[!th]
\begin{center}
\includegraphics[width=0.97\linewidth]{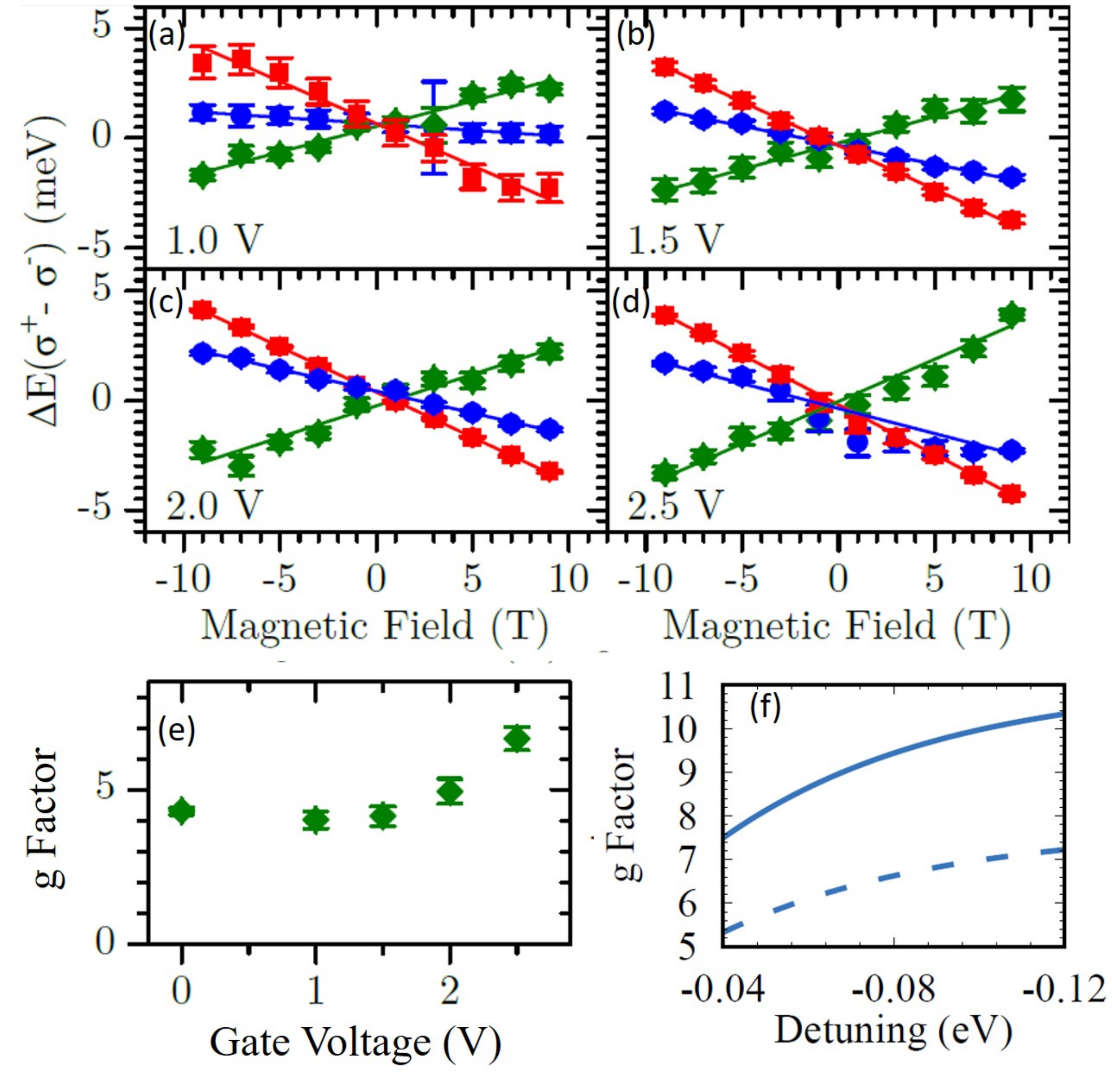}
\caption{a-d) Zeeman-splittings extracted at various bias conditions (1-2 V) for  X$^{\mathrm{*}}$ (red square) X$^{\mathrm{A}}$ (blue dot) and X$^{\mathrm{H}}$ green diamond. e) g-factor for X$^{\mathrm{H}}$ with respect to the gate voltage. f) Model curve for the g-factor for X$^{\mathrm{H}}$ with respect to the gate voltage, based on the phenomenological approach introduced in eqn. 9. The two different curves utilize different g-factors for the MoS$_2$ B-exciton as well as the interlayer exciton.  }
\label{Fig4}
\end{center}
\end{figure}

The hybridization of energy bands, giving rise to the X$^{\mathrm{H}}$ exciton resonance can be expected to sensible react on the precise mode energy (the detuning between the indicated transitions sketched in Fig. \ref{Fig1b} c), and thus can be manipulated by the externally applied electric field. We check this hypothesis by repeating to determine the Zeeman-splittings for the captured manyparticle complexes for a variety of applied gate voltages Fig. \ref{Fig4} a)-d) and plot the extracted Zeeman-splittings as a function of the gate voltage: The trion emission indeed only displays a modest modification of the extracted g-factor, and the g-factor of the exciton X$^{\mathrm{A}}$ fluctuates around  a value of 2. 
More importantly, the  hybrid exciton X$^{\mathrm{H}}$, which stands in the focus of this study, displays a progressive increase of the g-factor from 4.2 at 0 V  up to 7 at 2.5 V (Fig. \ref{Fig4}e).

The reason for this behavior is the combined action of tunnel coupling of the interlayer excitons to the B exciton (which leads to a mixing of the g-factors) and the tuning of the energetic position of the interlayer exciton. Without coupling, let us assume first, that a pure interlayer exciton would have a g-factor of about 12 and the B exciton of about -3. We note, however, that in particular in the MoS$_2$ system, strongly fluctuating g-factors are observed throughout the literature (those typically scatter between -2 and -4.5 for the A-exciton)  \cite{goryca2019revealing, cadiz2017excitonic}, hinting at an utmost sensible dependence on the precise bandstructure which is sensibly reacting on the surrounding. 
Now, coupling between the B-exciton and the interlayer exciton leads to redistribution of the g factors among them, which becomes efficient if the interlayer exciton and the B exciton are weakly detuned. The applied
electric field  increases the detuning for the studied lower interlayer exciton from the B exciton resulting in an increasing g factor.

This can be phenomenologically captured in a simplified model including only the B exciton $P^B$ and one interlayer exciton $P^I$. The simplified Bloch equations read\cite{katsch2019theory}
\begin{align}
(\hbar \omega - \tilde{E}^B)P^B = \Omega + T P^I \\
(\hbar \omega - \tilde{E}^I)P^I =T P^B,
\end{align}
with the energies $\tilde{E}^B = E^B + g_B B_z$ and $\tilde{E}^I = E^I + g_I B_z$ and $E^I = E^B + \Delta$, $\Delta<0$. The new energies of these coupled oscillator equations are given as
\begin{align}
&(\hbar \omega)_{\pm} = \frac{\tilde{E}^B + \tilde{E}^I}{2} \pm \sqrt{\frac{(\tilde{E}^B-\tilde{E}^I)^2}{4}+T^2} \\
&=E_B + \frac{\Delta}{2} + \frac{g_B+g_I}{2}B_z \pm \sqrt{\frac{((g_B-g_I)B_z - \Delta)^2)}{4} + T^2},
\end{align}
where the $+$ solution refers to the $B$ transition and the lower solution refers to the interlayer transition. The effective $g^{Eff}$ factor is given as
\begin{align}
g^{Eff}_{\pm}  = (\partial_{B_z} (\hbar \omega)_{\pm}\bigg \vert_{B_z=0} =\frac{g_B+g_I}{2} \mp \frac{(g_B-g_I)\Delta}{4\sqrt{T^2 + \frac{\Delta^2}{4}}} 
\end{align}

We can now plot the  effective $g$ factor as a function of the detuning $\Delta$ in figure \ref{Fig4} f). We notice, despite a slight offset, the model captures the observed phenomenological behavior of the bias-tuneable $g$ factor of X$^{\mathrm{H}}$. Indeed, the quantitative deviation most likely arises from uncertainties associated with the g-factor of the (uncoupled) interlayer transition and the B-exciton, which are not experimentally accessible in our study. However, we notice that our A-interlayer exciton indeed displays $g$ factor of $\approx$ 2, rather than the canonical value of 4 captured for most other TMDC A-exciton complexes. Let us thus consider reduced $g$ factors of the B-exciton as well as the interlayer exciton, rescaled by a factor of 1.5. We note, that the dashed line depicted figure \ref{Fig4} f) manages to quantitatively capture the experimentally observed tuning range of the interlayer exciton can now qua and the interlayer exciton $g$ factor.  \\

In conclusion, we have demonstrated the existence of an interlayer exciton in pristine bilayer MoS$_2$, which is characterized by a giant dipole moment in conjunction with a persistent oscillator strength. We have analyzed the behavior of this new excitonic species in an applied external electric and magnetic field, and found evidence for the mixing of K and K` valley states, contributing to the exciton. The valley character is tunable in an externally applied electric field, by modifying the effective interlayer- vs. intralayer character. 
Dipolar excitons with strong oscillator strength are of paramount importance for microcavity experiments in the regime of strong light-matter coupling. We anticipate, that the interlayer exciton in pristine bilayer MoS$_2$ can be utilized to generate strongly interacting, dipolar exciton polaritons based on atomically thin materials. \\

During the preparation of the manuscript, we became aware of a an independent, similar work by Leisgang et al.  arXiv:2002.02507 (2020).


\textbf{\textit{Acknowledgements --}} The Würzburg group gratefully acknowledges support by the state of Bavaria. C.S. acknowledges funding with the ERC project unLiMIt-2D (Project No: 679288) and the DFG with the priority programm SPP 2244 (DFG SCHN1376 14.1). Technical support by M. Emmerling and A. Wolf is acknowledged.  S.T. acknowledges support by the NSF (DMR-1955668 and DMR-1838443). We acknowledge fruitful discussions with our colleague Dominik Christiansen (TU Berlin). The TUB group was funded by the Deutsche Forschungsgemeinschaft via projects 182087777 in SFB 951 (project B12, M.S., and A.K.) and KN 427/11-1 (F.K. and A.K.). F.K. thanks the Berlin School of Optical Sciences and Quantum Technology.



\textbf{\textit{Methods --}} Sample fabrication: MoS$_2$ sheets were isolated from home-made flux zone grown MoS$_2$ crystals. The process started with 6N purity metal powder (Mo) and sulfur ingot pieces. Additional care was given to performing in house purification to eliminate all the other contamination (typically heavy metals and magnetic metals) to reach true 6N purity. Crystal growth was performed using flux zone growth technique without any transporting agents to capture high quality crystals. 
The bilayers were then isolated using the dry transfer technique introduce in by Castellanos-Gomez et al. \cite{castellanos2014deterministic}. we exfoliate our 2D materials on a transparent PolyDimethylSiloxane (PDMS) film. From this film, we transfer the 2D layers on a Si/SiO$_2$ substrate (with 100 nm SiO$_2$) with pre-patterned gold contact, that were initially deposited on the substrate via e-beam assisted evaporation.

\end{document}